\documentclass[superscriptaddress,twocolumn,prl]{revtex4-1}

\usepackage[latin9]{inputenc}
\usepackage{amsmath}
\usepackage{upgreek}
\usepackage{amstext}
\usepackage{graphicx}  
\usepackage{changes}
\usepackage{array}
\usepackage{subcaption}

\begin{document}
\title{An Improved Torsion Balance Test of the Equivalence Principle Towards the Sun}
\author{M.P. Ross}
\author{ E.A. Shaw}
\author{ C. Gettings}
\author{ S.K.Apple}
\author{I.A. Paulson}
\author{ J.H. Gundlach}
\affiliation{Center for Experimental Nuclear Physics and Astrophysics, University of Washington, Seattle, Washington 98195, USA}

\begin{abstract}

We search for violations of the Equivalence Principle towards the Sun using a rotating torsion balance apparatus. We set 95\%-confidence limits on violations with beryllium and aluminum test bodies of $\eta_{\odot, Be-Al} \leq 2.1 \times 10^{-13}$. These results are a factor of four improvement of previously reported results towards the Sun and a $\sim20\%$ improvement on previous torsion balance tests regardless of source.

\end{abstract}

\maketitle

\textit{Introduction} - The Equivalence Principle (EP) is a fundamental aspect of our geometric description of gravity (i.e. General Relativity). The EP states that gravitational acceleration is independent of the composition of objects. Deviations from the EP are traditionally parameterized with the Eotv\"os parameter \cite{Eotvos}:
\begin{equation}
\eta = \frac{a_1-a_2}{(a_1+a_2)/2}
\end{equation}
where $a_1$ and $a_2$ are the accelerations of two test particles. General Relativity predicts that $\eta=0$ for all choices of materials and attractors.

There have been many experimental tests of the EP over the history of physics with increasing precision and using a variety of methods. The systems deployed to test the EP have included torsion balance \cite{wagner2012torsion}, satellites \cite{Microscope}, and astronomical systems \cite{LLR}. The highest precision test to date comes from the MICROSCOPE mission \cite{Microscope} that set limits of $\eta_{\oplus, Ti-Pt}<5.5\times 10^{-15}$ for titanium and platinum towards the Earth. A significant number of laboratory EP tests use the Earth as the source mass. However, these tests do not probe interactions that do not couple to elements that make up the Earth.

Here, we describe a torsion balance test of the EP towards the Sun. The Sun as a source mass is of particular cosmological and astrophysical interest as it has a similar composition to most of the baryonic matter content of the universe, namely hydrogen and helium. This study is an extension of a search for long range interactions between normal matter and dark matter which is detailed in Ref. \cite{ross2025probing}. The data has also been searched for couplings to ultra-light vector dark matter that is detailed in Ref. \cite{ross2025search}.

\begin{figure}[!h]
    \includegraphics[width=0.5\textwidth]{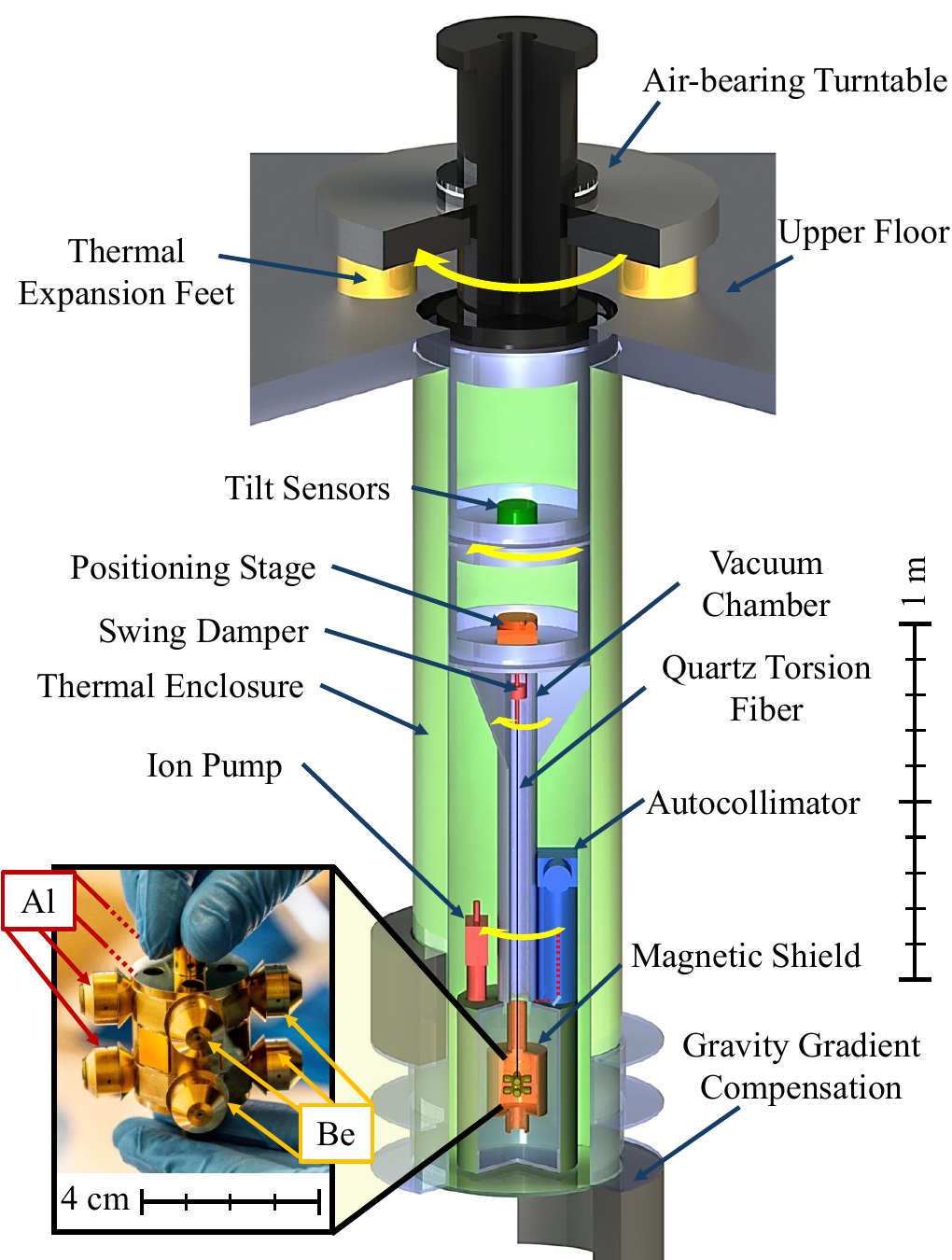}
    \caption{Schematic of the rotating torsion balance apparatus. The torsion pendulum, vacuum system, and angular readout are mounted on a rotating air-bearing turntable. The tilt of the apparatus is measured by a pair of co-rotating tilt sensors and controlled by thermal expansion feet that the turntable rests on. Thermal gradients and changes of the temperature, magnetic field, and gravity gradients are controlled, shielded, and compensated for, respectively. Reprinted from Ref. \cite{ross2025probing}.}
\label{apparatus}
\end{figure}

\textit{Apparatus} - We searched for differential accelerations towards the Sun with an upgraded version of our previously operated rotating torsion balance \cite{wagner2012torsion} located in an underground laboratory on the University of Washington campus. The upgrades are primarily in the form of a ultra-low loss fused silica torsion fiber \cite{shaw2022torsion} as well as modernization of the computing and control infrastructure. A schematic of the apparatus is shown in Fig. \ref{apparatus}. Further details of the apparatus can be found in Ref. \cite{ross2025probing, ross2025search}.

The torsion balance consisted of a pendulum with eight test bodies, four made of beryllium and four of aluminum, in a dipole orientation and was suspended from a 22-{\textmu}m thick, 1-m long fused quartz fiber \cite{shaw2022torsion}.

\begin{widetext}

\begin{figure}[!t]
\centering \includegraphics[width=\textwidth]{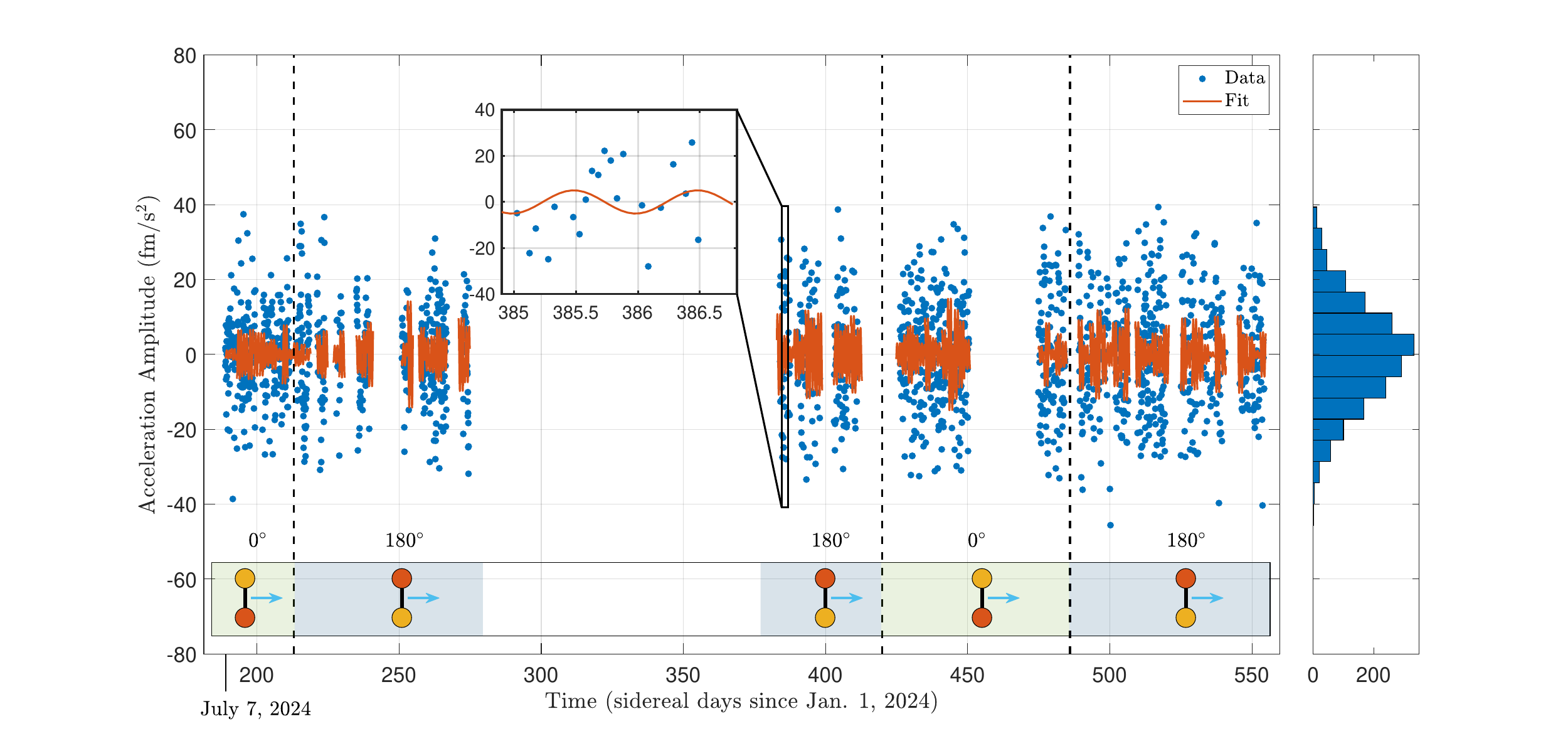}
\caption{Once per turntable revolution differential acceleration amplitudes ($\Delta a_{TT}$) along with the corresponding fits to the solar basis functions. Each two-day long cut was independently fit. Gaps in the data are due to local construction activity and hardware failures. The right pane shows the histogram of the acceleration amplitudes.}
\label{time} 
\end{figure}

\end{widetext}

The entire apparatus was held by an air-bearing turntable that rotated at 0.46 mHz. With each turntable rotation the test bodies on the pendulum were effectively interchanged in the lab frame. The angle of the pendulum relative to the vacuum chamber was measured by an autocollimator. The performance of the apparatus was limited by thermal noise \cite{thermal} below 1 mHz equating to $\sim$29 fm/$s^2$ at 0.46~mHz.

\textit{Analysis} - The experiment ran from July 7, 2024 to July 7, 2025. The pendulum was rotated $180^\circ$ on July 31, 2024, February 15, 2025, and May 1, 2025 to minimize systematic effects. The duty cycle of the apparatus was severely restricted due to nearby construction activity and an unexpected hardware failure. Although the data spans 366 days, only 186 days (48\%) were high quality.

The measured differential acceleration was split into sets that were two turntable rotations long. Each set was fit to a series of sinusoidal functions with frequencies of harmonics of the turntable frequency ($\omega_{TT}$, $2\omega_{TT}$, $3\omega_{TT}$, etc.) and harmonics of the torsional resonance ($\omega_0$, $2\omega_0$, $3\omega_0$, etc.). 
A series of once per turn-table revolution acceleration amplitudes was extracted as complex numbers ($\Delta a_{TT}$) to capture both the amplitude and phase of the differential acceleration.

The instrument experienced occasional transients due to earthquakes, construction activity, and other environmental disturbances. We discarded sets that did not fall under two criteria: misfit-squared less than four times thermal noise, and a $\Delta a_{TT}$ outside of the 95\% interval. Fig. \ref{time} shows the surviving once per turntable revolution acceleration amplitudes which are consistent with thermal noise. See Ref. \cite{ross2025probing} for details of statistical analysis.

\begin{figure}[!h]
\centering \includegraphics[width=0.5\textwidth]{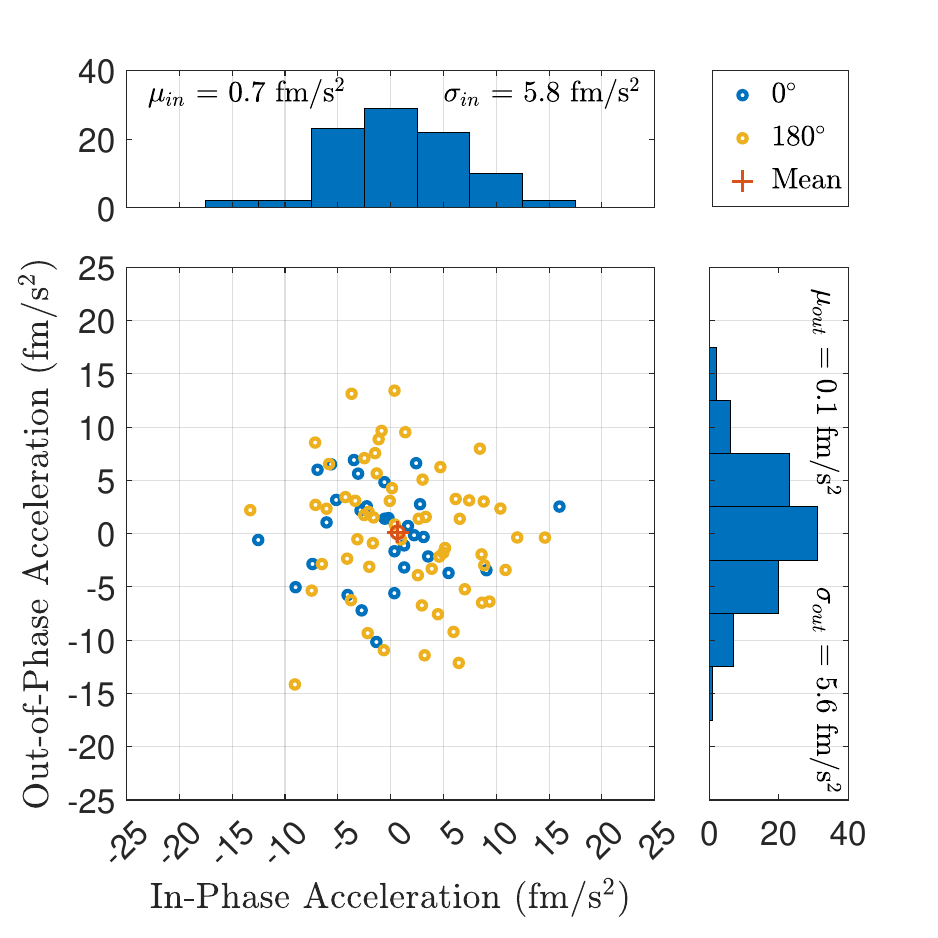}
\caption{Measured differential accelerations from each two-day long segments towards the Sun (in-phase) and the orthogonal direction (out-of-phase).}
\label{fits} 
\end{figure}

The $\Delta a_{TT}$ series was extracted in two-day long segments. Each segment was then fit to the solar basis functions. These functions consist of the location of the Sun projected into the horizontal plane of the lab and an orthogonally oriented function. These functions modulate once per day due to the rotation of the Earth and once per year due to the location of the lab at $\sim47^\circ$ latitude. The location of the Sun and the necessary coordinate transformations were computed using the {\tt Astropy}\cite{astropy:2022} libraries.

The series of measured differential accelerations toward the Sun is shown in Figure \ref{fits}. Measuring the out-of-phase component verifies the sensitivity of the experiment as the signal of interest would only appear as a non-zero in-phase component. We take the mean and corresponding uncertainty to give a final measurement of the differential acceleration towards the Sun:
\begin{equation}
    \Delta a_{\odot} = 0.66 \pm 0.61 \text{ fm/s}^2\quad (1\sigma)
\end{equation}

\textit{Results} - Given that the gravitational acceleration on Earth towards the Sun is $\sim$5.9 $\text{mm/s}^2$, the differential acceleration measurements imply a limit on EP-violation of:

\begin{equation}
    \eta_{\odot, Be-Al} \leq 2.1 \times 10^{-13}\quad (95\%\text{-confidence})
\end{equation}

These results are a factor of four improvement of previously reported results towards the Sun \cite{wagner2012torsion}. Additionally, this is a $\sim20\%$ improvement on previous torsion balance tests of the EP regardless of source or distance scale.

\textit{Conclusion} - The Equivalence Principle is foundational to our understanding of gravity. We have improved the reach of torsion balance tests of the EP and set stringent limits on the violations with solar matter. With future improvements, including faster turntable rate, improved turntable control, and interferometric angle readout, this apparatus may match or exceed the sensitivity of state of the art satellite tests.

\textit{Data Availability} - Raw data will be shared upon request. The fit amplitudes and analysis code can be found at \url{https://github.com/EotWash/EP-Analysis-Sun}.

\textit{Acknowledgements} - This work was supported by funding from the NSF under Awards PHY-1607385, PHY-1607391, PHY-1912380, and PHY-1912514.

\newpage

\bibliographystyle{unsrtnat}
\bibliography{main.bib}

\end{document}